\documentclass[11pt]{article}
\usepackage{moriond,epsfig}

\def\be{\begin{equation}}
\def\ee{\end{equation}}
\def\bea{\begin{eqnarray}}
\def\eea{\end{eqnarray}}

\begin{document}
\vspace*{4cm}

\title{Are the $\Theta^+(1540)$,
$\Xi^{--}(1860)$ and $D^{*-}p(3100)$ Pentaquarks or Heptaquarks?}

\author{P. Bicudo}

\address{
Dep. F\'{\i}sica and CFIF, Instituto Superior T\'ecnico, Av.
Rovisco Pais
1049-001 Lisboa, Portugal}

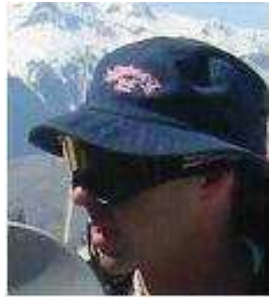
\maketitle

\abstracts{
We study the $\Theta^+$(1540) discovered at SPring-8. 
We apply Quark Model techniques, that explain with success the repulsive hard core of 
nucleon-nucleon and kaon-nucleon exotic scattering, and the short range attraction present 
in pion-nucleon and pion-pion non-exotic scattering. We find a K-N repulsion which 
excludes the $\Theta^+$ as a $K-N$ s-wave pentaquark. 
We explore the $\Theta^+$ as a crypto-heptaquark, equivalent to a $K-\pi-N$ borromean boundstate, 
with positive parity and total isospin I=0. 
The attraction is provided by the pion-nucleon and pion-kaon interactions. 
The other candidates to pentaquarks $\Xi^{--}(1860)$ , observed at NA49, and $D^{*-}p(3100)$, 
observed at H1, are also studied as linear molecular heptaquarks.
}

\section{Introduction}

\par
The  $uudd\bar s$ pentaquark $\Theta^+(1540)$ was discovered at LEPS
\cite{Nakano} 
and  DIANA
\cite{Barmin}. 
After the Jefferson Lab confirmation
\cite{Hicks}, 
it was observed in several different experiences, 
with a mass of 1540 $\pm$ 10 MeV and a decay width of  15 $\pm$ 15 MeV.
Recently the $ddss\bar u$ pentaquark $\Xi^{--}(1860)$ was observed at NA49
\cite{Alt}
and the $uudd \bar c$ pentaquark $D^{*-}p(3100)$ was observed at 	H1 
\cite{H1}.
These are extremely exciting states because they may
be the first exotic hadron to be discovered, with quantum numbers
that cannot be interpreted as a quark and an anti-quark meson or
as a three quark baryon. 
Exotic multiquarks are expected since the early works of Jaffe 
\cite{Jaffe,Strottman},
and some years ago Diakonov, Petrov and Polyakov
\cite{Diakonov}
applied skyrmions to a precise prediction of the $\Theta^+$. The
$\Xi^{--}(1860)$ and $D^{*-}p(3100)$ belong probably to the same family
of exotic flavour pentaquarks.

\par
We start in this talk by reviewing the Quark Model (QM) and the
Resonating Group Method (RGM) 
\cite{Wheeler}, 
which are adequate to study states where several quarks overlap. 
First we apply the RGM to show 
\cite{Bender,Bicudo,Barnes} 
that the exotic 
$N-K$ hard core s-wave interaction is repulsive, excluding 
the $\Theta^+$ as a bare s-wave pentaquark
$uddu\bar s$ state or a tightly bound s-wave $N - K$ narrow
resonance. However a $\pi$ could also be present in this system, in which 
case the binding energy would be of the order of $30 \ MeV$.
Moreover this state of seven quarks would have a positive parity,
and would have to decay to a p-wave $N - K$ system, which is
suppressed by angular momentum, thus explaining the narrow width
of the $\Theta^+$. We then put together the $\pi -N$, $\pi - K$ 
and $N-K$ interactions to show that the $\Theta^+$
is possibly a borromean \cite{borromean} three body s-wave
bound state of a $\pi$, a $N$ and a $K$, with positive parity and
total isospin $I=0$. Finally we extend the crypto-heptaquark picture
to flavour $SU(4)$ and study the $Xi^{--}$ and $D^{*-}p$ multiquarks.

\section{A Quark Model Criterion for Repulsion/attraction}

\par
We use a standard Quark Model Hamiltonian.
The Resonating Group Method is a convenient method to compute the 
energy of multiquarks and to study hadronic coupled channels.
The RGM was first used by Ribeiro 
\cite{Ribeiro} 
to explain the $N-N$ hard-core repulsion. 

\par
We compute the
matrix element
of the Hamiltonian
in an antisymmetrised
basis of hadrons,
\begin{equation}
V_{ meson \, A \atop baryon \, B} =
{ \langle \phi_B \, 
 \phi_A | (1+P_{AB})[ -( V_{13}+V_{23}+V_{14}+V_{24})  
 P_{13} +A_{23}+A_{14} ]| \phi_A \phi_B \rangle 
\over \langle \phi_B \, \phi_A | 
(1+P_{AB})(1-  P_{13}) | \phi_A \phi_B \rangle }
\ ,
\label{overlap kernel} 
\end{equation}
where the exchange operator $P_{14}$ produces the states 
colour-octet x colour-octet, expected in multiquarks,
and where $A_{23}$ is the quark-antiquark annihilation potential,
crucial to the chiral symmetry of the interaction
\cite{Bicudo3,Bicudo1}.

The exchange overlap results in a
separable potential, and we arrive at the criterion for the 
interaction of ground-state hadrons:
\\
- whenever the two interacting hadrons have
a common flavour, the repulsion is increased,
\\
- when the two interacting hadrons have a matching quark and 
antiquark the attraction is enhanced.

%
%
\begin{table}[t]
\caption{ This table summarises the parameters $\mu , \, v \,
,\alpha \, , \beta$ (in Fm$^{-1}$)
 and scattering lengths $a$ (in Fm) .
\label{scattering lengths}}
\begin{tabular}{|c|cccccc|}
\hline
channel                   & $\mu$  & $v_{th}$&$\alpha$ &$\beta$& $a_{th}$& $a _{exp}$ \\
\hline
$  K-N_{ I=0 }         $ & $ 1.65$ & $ 0.50$ & $ 3.2$ & $ 3.2$ & $-0.14$ & $ -0.13\pm 0.04 $
\cite{Barnes} \\
$  K-N_{ I=1 }         $ & $ 1.65$ & $ 1.75$ & $ 3.2$ & $ 3.2$ & $-0.30$ & $ -0.31\pm 0.01 $
\cite{Barnes} \\
$ \pi-N_{I={1\over 2}} $ & $ 0.61$ & $-0.73$ & $ 3.2$ & $ 11.4$ & $ 0.25$ & $  0.246\pm 0.007$
\cite{Itzykson} \\
$ \pi-N_{I={3\over 2}} $ & $ 0.61$ & $ 0.36$ & $ 3.2$ & $ 3.2$ & $-0.05$ & $ -0.127\pm 0.006$
\cite{Itzykson} \\
$ \pi-K_{I={1\over 2}} $ & $ 0.55$ & $-0.97$ & $ 3.2$ & $ 10.3$ & $ 0.35$ & $  0.27\pm 0.08 $
\cite{Nemenov} \\
$ \pi-K_{I={3\over 2}} $ & $ 0.55$ & $ 0.49$ & $ 3.2$ & $ 3.2$ & $-0.06$ & $ -0.13\pm 0.06 $
\cite{Nemenov} \\
\hline
\end{tabular}
\end{table}

\section{Why the $\Theta^+$ cannot be a simple $uudd \bar s$ or $K-N$ state}

Applying the criterion to the S=1, I=0 pentaquark, arranged in the color
singlet clusters $uud-d \bar s$  or $ ddu-u \bar s$
we find repulsion! 
Indeed we are able to reproduce the repulsive K-N exotic s-wave phase shifts, 
which have been understood long ago
\cite{Bender,Bicudo,Barnes}.
Moreover all other $uudd \bar s$ systems are even more repulsive or unstable. 
Because we checked all our only approximations, say using a variational 
method, and neglecting the meson exchange interactions, we estimate that 
something even more exotic is probably occuring!

Suppose that a $q-\bar q$ pair is added to the system.
Then the new system may bind. Moreover the heptaquark had a different parity 
and therefore it is an independent system (a chiral partner).
Here we propose that the $\Theta^+$ is in fact a heptaquark with the strong 
overlap of a $K-\pi-N$, where the $\pi$ is bound by the I=1/2 $\pi-K$ and 
$\pi-N$ attractive interactions.

%
%
%
\begin{figure}[t]
\begin{picture}(400,160)(0,0)
\put(0,150){\bf (a)}
\put(0,50){
\begin{picture}(200,70)(0,0)
\put(20,5){
\begin{picture}(120,70)(0,0)
\put(20,0){
\begin{picture}(100,100)(0,0)
\put(0,0){\oval(30,30)}
\put(0,5){$N$}
\put(-12,-8){$I={1\over 2}$}
\put(40,55){\oval(30,30)}
\put(35,43){$\pi$}
\put(30,58){$I=1$}
\put(80,0){\oval(30,30)}
\put(70,5){$K$}
\put(68,-8){$I={1\over 2}$}
\end{picture}}
\put(20,0){
\begin{picture}(20,100)(0,0)
\put(40,10){\oval(90,14)}
\put(25,5){$I=1$}
\put(41,40){\oval(20,20)[tl]}
\put(31,40){\line(1,-1){40}}
\put(41,50){\line(1,-1){40}}
\put(71,10){\oval(20,20)[br]}
\put(43,30){$I$}
\put(46,27){$=$}
\put(54,25){$1 \over 2$}
\put(6,10){\oval(20,20)[bl]}
\put(-4,10){\line(1,1){40}}
\put(5,-1){\line(1,1){40}}
\put(36,40){\oval(20,20)[tr]}
\put(15,21){$I$}
\put(20,26){$=$}
\put(26,29){$1 \over 2$}
\end{picture}}
\put(20,0){
\begin{picture}(20,100)(0,0)
\put(-20,60){Total}
\put(-20,50){$I=0$}
\end{picture}}
\end{picture}}
\end{picture}
}
\put(200,150){\bf (b)}
\put(200,0){
\begin{picture}(245,180)(0,0)
\put(25,40){
\begin{picture}(220,60)(0,0)
\put(0,0){
\begin{picture}(100,100)(0,0)
\put(0,0){\line(1,0){60}}
\put(0,0){\line(2,1){40}}
\put(60,0){\line(-1,1){20}}
\put(28,25){$dds$}
\put(0,-10){$ddss\bar u$}
\put(-30,0){1,86}
\put(-30,-10){GeV}
\put(-16,20){1,38 GeV}
\end{picture}}
\put(60,0){
\begin{picture}(100,100)(0,0)
\put(0,0){\line(1,0){60}}
\put(0,0){\line(2,1){40}}
\put(60,0){\line(-1,1){20}}
\put(20,25){$uds$}
\put(-5,-10){$dss$}
\end{picture}}
\put(120,0){
\begin{picture}(100,100)(0,0)
\put(0,0){\line(1,0){60}}
\put(0,0){\line(2,1){40}}
\put(60,0){\line(-1,1){20}}
\put(20,25){$uus$}
\put(-5,-10){$uss$}
\put(55,-10){$usss \bar d$}
\end{picture}}
\put(80,80){
\begin{picture}(100,100)(0,0)
\put(0,0){\line(1,0){60}}
\put(0,0){\line(3,2){40}}
\put(60,0){\line(-3,4){20}}
\put(40,30){$uudd\bar s$}
\put(-15,5){$udd$}
\put(65,0){$uud$}
\put(-60,0){1,86 GeV}
\put(-15,25){1,54 GeV}
\end{picture}}
\put(40,20){
\begin{picture}(100,100)(0,0)
\put(0,0){\line(1,0){60}}
\put(0,0){\line(2,3){40}}
\put(60,0){\line(-1,3){20}}
\end{picture}}
\put(100,20){
\begin{picture}(100,100)(0,0)
\put(0,0){\line(1,0){60}}
\put(0,0){\line(2,3){40}}
\put(60,0){\line(-1,3){20}}
\end{picture}}
\put(110,4){
\begin{picture}(100,100)(0,0)
\put(-5,0){$0$}
\put(0,0){\vector(1,0){90}}
\put(60,-2){\line(0,1){4}}
\put(54,1.5){$1$}
\put(79,5){$m_I$}
\put(-11,-45){\vector(1,4){22.5}}
\put(-12,-40){\line(1,0){4}}
\put(8,40){\line(1,0){4}}
\put(-7,-43){$-1$}
\put(13,37){$1$}
\put(-20,-45){$S$}
\put(0,0){\vector(0,1){122}}
\put(-2,41){\line(1,0){4}}
\put(-2,82){\line(1,0){4}}
\put(-7,40){$1$}
\put(-7,80){$2$}
\put(10,120){$M$}
\put(25,120){$[GeV]$}
\end{picture}}
\end{picture}}
\end{picture}
}\end{picture}
\caption{ We show in {\bf (a)} the isospin couplings in the $\Theta^+$.
In {\bf (b)} we exhibit, in a three dimensional strangeness-flavour-mass plot,
the expected masses of the exotic anti-decuplet.
}
\label{isospin}
\end{figure}
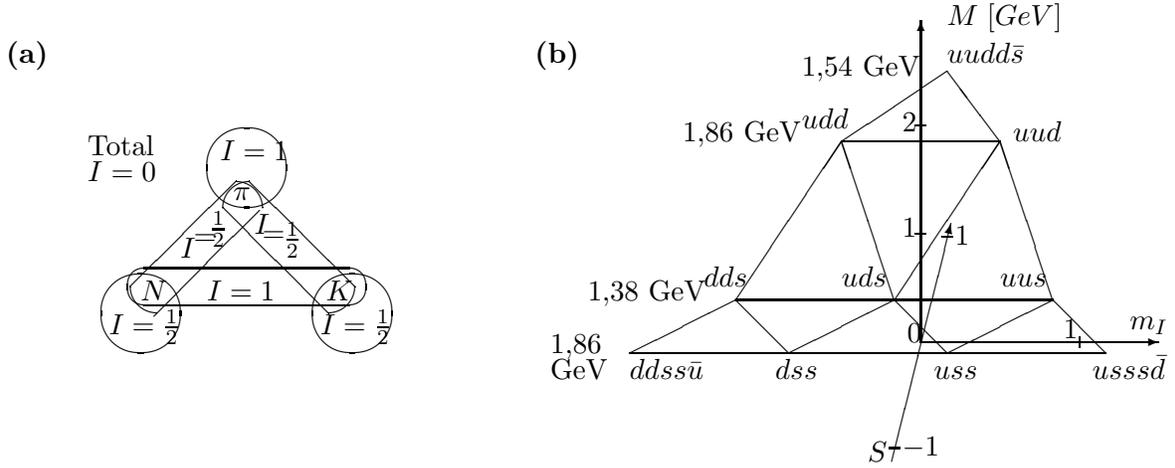

\section{the $K-N$, $\pi-K$, $\pi-N$ and $K-\pi-N$ systems}

\par
We now investigate the borromean \cite{borromean} binding of the
exotic $\Theta^+$ constituted by a $N$, $K$ and $\pi$ triplet. 
We arrive at the separable potentials for the different two-body potentials 
\cite{Bicudo,Bicudo1},
\begin{eqnarray}
V_{K-N}&=& \frac{2-\frac{4}{3}  \vec \tau_A \cdot \vec
\tau_B}{\frac{5}{4}+\frac{1}{3}  \vec \tau_A \cdot \vec \tau_B}
\frac{(M_\Delta-M_N )}{3} \left( 2 \sqrt{\pi} \over \alpha \right)^3 
e^{- \frac{{p_\lambda}^2}{2\beta^2
}} \int \frac{d^3 p'_\lambda}{(2 \pi)^3} \, e^{- \frac{
{p'_\lambda}^2}{2 \beta^2}}
\nonumber \\
V_{\pi-N}&=& \frac{2}{9}( 2M_N - M_\Delta ) \,  \vec \tau_A \cdot
\vec \tau_B \, {\cal N_\alpha}^{-2} \ ,
\nonumber \\
V_{\pi-K}&=& \frac{8}{27}(2 M_N - M_\Delta) \, \vec \tau_A \cdot
\vec \tau_B \, {\cal N_\alpha}^{-2} \ , \label{zero p}
\end{eqnarray}
where $\vec \tau$ are the isospin matrices. 

\par
Because the pion is quite light we start by computing the 
pion energy in an adiabatic K-N system. 
Our parameter set, tested in 2-body channels, 
is presented in Table \ref{scattering lengths}.
The only favourable flavour combination is
shown in Fig. \ref{isospin} {\bf (a)}. Indeed we get quite a 
bound pion, but it only binds at very short $K-N$ distances.
However when we remove the adiabaticity, by allowing 
the K and N to move in the pion field, we find that the 
pion attraction overcomes the $K-N$ repulsion but not yet the
the $K-N$ kinetic energy. Other effects may further increase
attraction. We are planning to include full three-body Fadeev
equations, the coupling to the $K-N$ p-wave channel and the
short-range two-pion-exchange-interaction.

\section{ $SU(4)$ flavour: the $\bar K -N-\bar K $ and anti-charmed systems}

\par
Extending the pentaquark and the molecular heptaquark picture to the full SU(3) 
anti-decuplet we arrive at the picture shown in Fig. \ref{isospin} {\bf (b)}, where,
\\
-The $\Xi^{--}(1860)$  cannot be a $ddss \bar u$ pentaquark because it would
suffer from repulsion.
\\
- Adding a $q-\bar q$ pair we arrive at a $I=1/2$ $\bar K-N- \bar K$ linear molecule
where the the $\bar K-N$ system has isospin I=1, and it is an attractive system. 
We find that the $\bar K-N- \bar K$ molecule is bound, although we are 
not yet able to arrive at a binding energy of  - 60 MeV. 
\\
- Then the $I=1/2$ elements of the exotic anti-decuplet are $K-\bar K-N$ molecules.
\\
- Only the I=1 elements are pentaquarks, or equivalently overlapping $\bar K-N$ 
systems.

In what concerns anti-charmed pentaquarks like the very recently observed $D^{*-}p$, 
or anti-bottomed ones, this extends the anti-decuplet to flavour $SU(4)$ or $SU(5)$. 
Anti-charmed pentaquarks were predicted by many authors, replacing the s by a c.
Again the pentaquark $uudd \bar c$ is unbound, and we are researching the possible 
$D(D^*)-\pi-N$ molecular heptaquarks.

\section{ Conclusion }

We conclude that the $\Theta^+1540)$, $\Xi^{--}(1860)$ and $D^{*-}p(3100)$ hadrons very 
recently discovered cannot really be s-wave pentaquarks. 

- We also find that they may be a heptaquark states, with two repelled 
$K$ and $N$ clusters bound third $\pi$  cluster. 

- More effects need to be included, say exact Fadeev equations, the K-N p-wave coupled 
channel, and medium range interactions. 

- This is a difficult subject with the interplay of many effects. The theoretical models 
should not just explain the pentaquarks, they should also comprehend other hadrons. 

\section*{Acknowledgments} 

Most of the work presented here was done in collaboration with Gon\c{c}alo Marques
\cite{Bicudo2}.

\end{document}